# Efficient Wu-Manber Pattern Matching Hardware for Intrusion and Malware Detection

Monther Aldwairi
College of Technological Innovation
Zayed University
Abu Dhabi 144534, UAE
monther.aldwairi@zu.ac.ae

Yahya Flaifel, Khaldoon Mhaidat
Faculty of Computer and Information Technology
Jordan University of Science and Technology
Irbid 22110, Jordan
yahya.flaifel@csb.gov.jo, mhaidat@just.edu.jo

*Abstract*—Network intrusion detection systems and antivirus software are essential in detecting malicious network traffic and attacks such as denial-of-service and malwares. Each attack, worm or virus has its own distinctive signature. Signature-based intrusion detection and antivirus systems depend on pattern matching to look for possible attack signatures. Pattern matching is a very complex task, which requires a lot of time, memory and computing resources. Software-based intrusion detection is not fast enough to match high network speeds and the increasing number of attacks. In this paper, we propose special purpose hardware for Wu-Manber pattern matching algorithm. FPGAs form an excellent choice because of their massively parallel structure, reprogrammable logic and memory resources. The hardware is designed in Verilog and implemented using Xilinx ISE. For evaluation, we dope network traffic traces collected using Wireshark with 2500 signatures from the ClamAV virus definitions database. Experimental results show high speed that reaches up to 216 Mbps. In addition, we evaluate time, device usage, and power consumption.

## I. INTRODUCTION

Nowadays, information technology and computers play an essential role in all aspects of our lives. A gargantuan amount of digital data is produced in social, personal, multimedia, government, business and scientific domains. Therefore, the storage, maintenance, analytics and safeguarding of such data against intrusions had become an important research topic. Intrusion detection systems and antivirus software help protect the data at the edge of the network and end device, respectively.

Network intrusion detection system (NIDS) is essential in detecting malicious network traffic. Based on the approach, there are two categories: anomaly and signature-based. Anomaly works by detecting any behavior deviating from the baseline. The baseline is a profile capturing normal activities of a computer system or network [1].

Anomaly-based have self-learning capabilities. The system administrator trains the system to recognize the baseline. The normal state of the network's traffic is defined using parameters such as: payload, packet size, and protocol. After training finishes, the anomaly-based system monitors network and compares its state with the normal baseline to detect anomalies. This approach is effective in capturing new attacks and behaviors, however it generates many false positives and negatives [2].

Signature-based NIDS, also called misuse-based, relies on a set of rules to define suspicious activities. Those rules are applied to all incoming and outgoing traffic. Rules define several parameters that identify the suspicious activity such as the packet type, a signature string to match against the traffic, a location where the signature might be, and a type of action to take if the packet is deemed malicious. Rules come in several forms depending on the particular system [3]. In summary, signature-based NIDS simply searches the data for exact signatures. The existence of those signatures means with 100% certainty that the packet or data contains an attack or malware. Signature-based NIDS remains the most commonly used up to date because of their accuracy and superior speed [4].

Both anomaly or signature systems need to monitor the network packets in real-time, either by comparing patterns or by finding out-of-ordinary behavior [5]. Signature-based mostly depends on a pattern matching algorithm to perform the core matching process. Pattern matching is a very complex and time-consuming task, because it happens at the ever-increasing network speed and against the increasing number of malware or attack definitions. Software-based NIDS running on general purpose hardware are not fast enough to deal with the high network speeds and the increasing number of attacks. Therefore, NIDS require dedicated hardware that is designed specifically for this critical and challenging task [6].

Field programmable gate arrays (FPGAs) are arrays of programmable logic that can be configured to perform any computational task. They are excellent candidate for fast matching against large patterns because they are a large array structure of configurable logic, arithmetic, and memory blocks that can work on many tasks at the same time. FPGAs are re-programmable, which makes it easier to accommodate new attack and malware signatures. Moreover, hardware upgrades and bug fixes can be as easy as fixing a software [7].

In this paper, we propose a hardware specific Wu-Manber algorithm. We implement it on FPGA to speed up NIDS/antivirus. The rest of this paper is organized as follows. Section 2 explains pattern matching and Wu-Manber algorithm. Section 3 provides a critical analysis of the related work. The proposed hardware design is explained in Section 4. Section 5 covers evaluation and analysis of bit rate, time, device usage, and power.

## II. BACKGROUND

This section sheds more light on Snort NIDS and ClamAV antivirus. It also describes pattern matching and explains the inner workings of Wu-Manber.

### A. Snort Intrusion Detection System

Signature-based NIDSs were developed to search network traffic for predefined attack patterns. Snort is popular open source, NIDS acquired by CISCO [8]. Snort comes with all sorts of built-in rule files. The rule files are ASCII coded text files with each line representing a rule. Figure 1 shows Snort rule format and gives a real example. Snort rule with *sid 2002383* alerts the system administrator of a potential FTP brute-force login scan attempt. If an outgoing *tcp* packet from port 21 (*ftp*) contains the signature "530", *ftp* code for failed login.

### B. ClamAV Anivirus

Antivirus software protects networks and clients from malwares. ClamAV is a very popular open source antivirus available under General Public License (GPL) [9-11]. It has its own proprietary format for virus definitions: ClamAV Virus Database (.CVD) files. ClamAV comes with two files: main.cvd and daily.cvd, that are updated automatically. Those .CVD files have a text based 512 bytes header with colon separated fields. The signatures can be opened and extracted using the provided command line sigtool.

### C. Pattern Matching

Pattern matching is one of the most fundamental operations in computer transactions from search to pattern recognition and bioinformatics [12]. It is an essential and critical part of many applications including search engines, anti-malware, antivirus, spam filters, and intrusion detection and prevention systems. Therefore, there is a great need for fast and efficient pattern matching [13].

Pattern matching algorithms can be divided into two main categories depending on the approach: exact and approximate pattern matching. In exact pattern matching, all characters of the pattern must exist in the search string to declare a successful match. This type of algorithms is more prevalent in anti-malware, antiviruses and intrusion detection systems. In approximate pattern matching, a match is successful if a string similar to the given pattern is found. That is inexact match, where the pattern found differs by *k* characters, called edit distance, from the required signature. This type of algorithms is used in SPAM filtering systems and information retrieval [14].

There exists many pattern matching algorithms, and several of them have been used by intrusion detection systems. The most popular algorithms include: Knuth-Morris-Pratt (KMP) [15], Boyer-Moore (BM) [16], Aho-Corasick (AC) [17], and Wu-Manber (WM) [18]. Each has advantages and disadvantages making them either attractive or less attractive choice for NIDS. KMP and BM fall into the single pattern matching category. Hence, we must examine the packet once for each attack signature, which is highly inefficient for NIDS. On the other hand, AC and WM are multi-pattern matching algorithms, in that they match all signatures against all packets in one sweep. They add a preprocessing stage, where all signatures are built into either hash tables in WM or a trie in AC. Both have been used by Snort [8]. However, WM has proven to be more efficient, for longer signatures, than AC. Additionally, adding new signatures in WM is much easier than AC, which requires rebuilding the finite state machine [19].

### D. Wu-Manber Algorithm

Wu-Manber algorithm is based on the basic principles of Boyer-Moore, but uses a block of size *B* instead of one character [20]. WM builds two main tables: shift and hash. The shift table contains the number of characters to skip forward in the case of a mismatch based on BM bad character heuristic. In the case of a probable match the hash table is searched. WM calculates the hash value of the suffix block of characters from the signatures. It is recommended by Wu and Manber that WM block size, *B*, be 2 or 3. The size of matching window is dictated by the length of the shortest signature.

WM has preprocessing and search stages [21]. Before any search takes place, the signatures must be preprocessed.

#### 1) Preprocessing stage

Preprocessing has two steps: first, the algorithm determines the size of the matching window, and second it builds the shift and hash tables. WM stores the shift forward distance of the character block in the shift table. Furthermore, it stores the entry of the linked list that contains all patterns with the same suffix in the hash table [22]. For example, if we have a pattern set $P = \{P_1, P_2, ..., P_k\}$ and text to search $T$, of length *n*, and a match window of size *m*, and $B=2$, then the shift table is constructed as follows. The shift value for any block *x* located within the current matching window is:

$$SHIFT[x] = \begin{cases} m\text{-}B + 1, \text{if } x \text{ does not occure in any pattern} \\ m\text{-}q \quad , otherwise \end{cases}$$

Where, *q* is the rightmost place *x* occurred in any of the patterns. If the shift for a block is zero then all patterns containing *x* are inserted as a linked list in the hash table.

```
[Action][Protocol][SourceIP/Mask][SourcePort]→
[DestinationIP/Mask][DestinationPort][Options]
```

```
alert tcp $HOME_NET 21 → $EXTERNAL_NET any
(msg:"ET SCAN Potential FTP Brute-Force attempt";
content:"530 "; pcre:"/530\s+(Login|User|Failed|
       Not)/smi"; sid:2002383; rev:10;
```

Fig. 1. Snort rule format and example

*2) Search stage*

To search through a packet, a sliding window is shifted forward. Keeping in mind that the WM block size, $B=2$, WM search works as follows. WM hashes the last two characters of the current sliding window. The hash value is the index to access the shift table. If the shift was $>0$, the sliding window is shifted forward and the process repeats. If the shift was zero, this means that the examined block is a suffix of at least one of the patterns. In this case, WM checks the hash table, using the same index used to access the shift table, to search the linked list for the patterns with the same suffix. Those patterns are then matched one by one. After the match process finishes, the search pointer is incremented by one if no match is detected, or incremented by the size of detected pattern and the process repeats.

### III. RELATED WORK

There had been a lot of efforts to speedup pattern matching in hardware dating back to the 1990's [23]. Cho, Navab and Mangione [3] proposed a parallel rule-based inspection firewall that processed each rule separately. Data packet is passed to the parallel rule units that compare headers against predefined header data. If a match is found, it passed the payload data to the content pattern match unit. The content pattern match unit had four 8-bit comparators to increase the throughput. It was tested using Snort signatures and achieved over 2.88 Gbps on an Altera EP20K with operating frequency of 90 MHz. Ten logic cells are required per search pattern.

One of our most recent efforts was a GPU-based accelerator for Myers algorithm. It parallelized the bit-vector approximate matching algorithm, on a multi-core CPU under the MapReduce framework. MAPCG achieved more than 4.5 times speedup over the serial version for normal network traffic. However, the memory overhead was high reaching over 2 GB [4].

Fang, Katz, and Lakshman [24] implemented an IDS scheme that deeply analyzed the intruder's signatures and categorized them into: simple, complex and correlated patterns. The authors use TCAMs to store patterns due its benefits in compressing contents size and implementing wildcards. The scheme consisted of static RAM memory used as partial hit list (PHL). According to the simulation, this scheme operated at 2 Gbps when matching with 240kB TCAM containing 1768 ClamAV signatures. The disadvantage is when the intruder intentionally sends packets, that PHL access rate will be very high, which affected the system throughput.

Tan and Sherwood [25] proposed a high throughput string matching architecture. They modified AC using bit-split parallelism by creating eight parallel smaller FSMs, one per bit. One of the advantages was that the signatures may be added without interrupting operation. They claimed they could build a 10 times faster system. However, they cannot expand beyond eight parallel instances.

Alicherry, Muthiprasanna and Kumar [26] proposed a high-speed pattern matching system that extends AC algorithm to make it capable of processing multiple characters at a time. In addition, they used TCAM to construct compressed AC DFA. The compressed DFA had multiple character transitions to achieve multi-gigabit rate search speed. Salour and Su [27] on the other hand proposed a dynamic two-layer signature based NIDS with unequal databases. They divided the signatures into two databases: small one containing common attacks, and the other database containing all remaining signatures. The system automatically decided, which database it should use: the small efficient database to improve the detection speed, or the complementary less than ideal large database.

Kharbutli, Aldwairi, and Mughrabi [28] proposed three novel Wu-Manber parallelism approaches: shared position, trace distribution and a combination of both. In shared position algorithm, several parallel scanning windows each working on different processors. All scanning windows shared the highest scanned position in the string that one of the scanning windows resided on. On the other hand, trace distribution algorithm, is regular data parallelism that distributed the network traces over multiple processing units. The disadvantage of this algorithm is that it needed prior knowledge of the dataset size. The third algorithm combined the two solutions by using multiple scanning windows over segmented dataset. They used Snort signatures and showed about 2 times speedup relative to serial implementation of Wu-Manber.

Themistoklis, Charalampos and Konstantinos [29] implemented WM algorithm using OpenCL Framework. The goal was to increase the performance of the GPUs when locating nucleotides and amino acid sequence patterns within biological databases. The experimental results showed a significant speedup of about 31 times for the best case, and about 2 times for the worst case. Maier [30] implemented Rabin-Karp algorithm on FPGAs. In preprocessing stage, the system hashed all patterns, which supported finite number of different patterns lengths because separate text hash must be calculated for every pattern length. They used Xilinx Virtex II and IV FPGAs and were able only to support 300 and 1500 patterns respectively. Unfortunately, those patterns are not enough and Rabin-Karp algorithm allows false positives.

Rafla and Gauba [31] introduced FSM-based hardware with embedded softcore processor MicroBlaze. The FSM can be reconfigured on-the-fly by altering the memory contents without performing place-and-route process. KMP [15], which is a single pattern matching algorithm, was used! The MicroBlaze accepted new patterns from the host system via HyperTerminal and a C program computed the prefix array. Then the MicroBlaze processor reconfigured the FSM by updating state transition and output vector tables according to the new prefix array. The results showed increased performance, however, the number of blocks needed to match a pattern within a text is equal to the length of the text plus the length of the pattern.

### IV. HARDWARE FOR PATTERN MATCHING

Wu-Manber algorithm is hash table based, which makes it suitable for FPGAs. The input text enters the system through the LAN interface module (LI). The Patterns Buffer (PB) module passes the patterns to the preprocessing stage to obtain the hash and shift values for the pattern pieces. The Pattern Shifter module (PS) finishes the preprocessing by creating the shift and hash tables. The hardware is now ready to receive the packets through the data stream to match against the patterns. The search is handled by the Pattern Matching module (PM), which matches input stream with the programmed patterns in the PB. The controller module (CM) controls all data paths, multiplexers, registers and control signals of all modules.

### A. Pattern Matching Module

PThe PM module performs the actual matching by taking input stream into buffer called Shift Buffer. It then searches for possible edges by tracking the two bytes residing at the locations away from the end of the buffer by ML and ML–1 as shown by Figure 2. Next, it retrieves the shift value of the two bytes' block from the shift table. If an edge is detected, the PM retrieves the desired pattern from patterns buffer and stores it in the match buffer to start matching between the two buffers.

Because the shift table can grow very large, the shift value retrieval for each block is expensive. A Bloom filter (BF) is used. It has a hash function circuit to compute a hash for each byte-pair. In preprocessing, when the PS computes a shift value for a byte-pair, the BF computes the hash value of the pair and sets the corresponding bit in the BF vector register. In matching, the BF is used to determine if the byte-pair does not have a shift in the table. If the BF bit is 1 then, the shift value is read from the table. Otherwise, the maximum shift is used.

### B. Pattern Buffer

The PB holds the patterns to match against, segmented into 64-bit words. The last word of the pattern is padded with zeros. The internal buffer is 65-bit wide, with 1 bit flag used to mark the last word. The internal buffer is addressed by the 32-bit Address Register (AD). All patterns with same byte-pair ending are stored next to each other in the buffer. This will make the PB each segment holds many patterns with same suffix [32]

### C. Hash Table

Hash table is a memory with 128k words, 32-bit each. It holds the start and end addresses of the PB segment holding the patterns for the same byte-pair suffix. The first half is for end addresses and the second is for start addresses. For example, HASH[1AA] is the start address of the PB segment that contains patterns that have patterns substrings of length ML and ending with "AA". HASH[0AA] is the end address of the same segment. At the preprocessing stage, when reaching the pattern's end, the address of the first word of the pattern set with the same ending is stored in the buffer, and the number of locations that the pattern holds in the PB is passed to HT through the 8-bit Bus1. Using this information, the HT computes the end address of the segment. When there is another pattern ending with same byte-pair, the HT uses Hash Count to compute the new end address of the pattern segment. The HT is an FSM with 7 states as shown by Figure. 3.

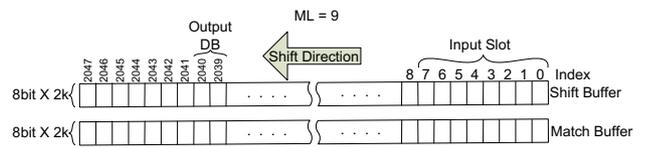

Fig. 2. PM buffers.

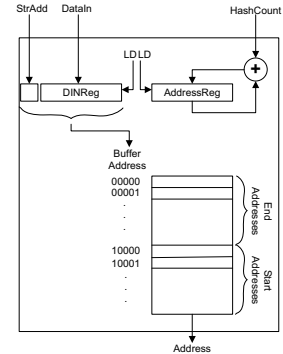

Fig. 3. Hash table module.

### D. Shift Table

The Shift Table (ST) consists of 64k byte memory that is addressed by byte-pair 16-bit Bus2. It holds the shift values of the corresponding byte-pair. The "write" input control is used to program the table during preprocessing, where shift values from Bus1 are written to locations addressed by Bus2. If the "write" signal is not asserted, the ST is in the reading mode. The "DataReady" and "WriteDone" flags indicate the completion of the read or write operations.

## V. EXPERIMENTAL RESULTS

Below we present the test datasets and a full evaluation in terms of device utilization, timing, power and bitrates.

### A. Datasets Collection and Processing

Two types of data were collected: the signatures from ClamAV [10] database, and the network traffic traces collected using Wireshark network protocol analyzer [33]. To extract the signatures from ClamAV CVD files we skipped the 512-byte header. There are several types of signatures in CVD, each with its own format. Excel was used to clean the signatures, extract the patterns and prepare them to be stored in the PB module. Each pattern was divided into 64-bit pieces, with 1-bit "PattFlag=0" to mark the last piece.

Excel was used to extract packet traces from the Wireshark output files. The cleaned trace file contains traffic frames, each on one line. This data must be segmented to 64-bit words to store them in the LI buffer. We develop a utility (FileSplit) to split the resultant string into lines. We collected traces of different traffic types: Mixed1&2 contain normal traffic such as browsing, video streaming, and chatting. In addition, we collected file download1 & 2 and RemoteControl1 & 2 traffic.

To generate malicious traffic, we mix random sets of patterns (Set1, 2, 3 and 4) into the traces randomly to generate doped trace files shown in table 1. The doping percentage is the number of lines with patterns over the total number of lines in the doped file.

*B. Synthesis Results*

The design was synthesized for Virtex 5 using Xilinx ISE Project Navigator 14.3 and simulated using ISim. Each module was synthesized alone to determine the device utilization and the maximum speed that can be reached.

*1) Device utilization*

Pattern matching module was the largest in terms of device utilization with 14,000 slices from the target device.

*2) Timing analysis*

We translate, map, place and route each module several times until timing constraints are met. Table 2 shows minimum delay Post-Place and Route timing reports (PPAR). The maximum speed for each module is shown by Figure 4. Obviously, PM was the slowest module with maximum frequency of 239MHz.

*3) Power analysis*

The power analysis using xPower Analyzer shows 2.5W consumption for all main modules, except for the controller, which consumed 1.5W because it has no large storage elements. The total power of the system is 47W.

TABLE I. TRAFFIC TRACES AND DOPING PERCENTAGE

| File | Size (kB) | Doping % |
|---|---|---|
| Set1Mixed1 | 238 | 39.02% |
| Set2Mixed1 | 404 | 63.98% |
| Set3Mixed1 | 147 | 1.66% |
| Set4Mixed1 | 163 | 11.71% |
| Set1Mixed2 | 238 | 39.02% |
| Set2Mixed2 | 404 | 63.98% |
| Set3Mixed2 | 147 | 1.66% |
| Set4Mixed2 | 163 | 11.71% |
| Set1FileDownload1 | 238 | 39.02% |
| Set2FileDownload1 | 404 | 63.98% |
| Set3FileDownload1 | 147 | 1.66% |
| Set4FileDownload1 | 163 | 11.71% |
| Set1FileDownload2 | 238 | 39.02% |
| Set2FileDownload2 | 404 | 63.98% |
| Set3FileDownload2 | 147 | 1.66% |
| Set4FileDownload2 | 163 | 11.71% |
| Set1RemoteControl1 | 238 | 39.02% |
| Set2RemoteControl1 | 404 | 63.98% |
| Set3RemoteControl1 | 147 | 1.66% |
| Set4RemoteControl1 | 163 | 11.71% |
| Set1RemoteControl2 | 238 | 39.02% |
| Set2RemoteControl2 | 404 | 63.98% |
| Set3RemoteControl2 | 147 | 1.66% |
| Set4RemoteControl2 | 163 | 11.71% |
| Average | | |

TABLE II. TIMING RESULTS

| Module | Min Period | Max Frequency (MHz) |
|---|---|---|
| Controller | 2.747ns | 364.033 |
| PM | 4.184ns | 239.005 |
| PS | 2.863ns | 349.284 |
| ST | 1.999ns | 500.250 |
| HT | 1.999ns | 500.250 |
| PB | 1.999ns | 500.250 |
| LI | 3.103ns | 322.269 |

TABLE III. BIT RATE VALUES AND SIMULATION TIME

| Signatures | Set1 | | Set2 | | Set3 | | Set4 | |
|---|---|---|---|---|---|---|---|---|
| File | Processing Time ms | Bit Rate Mbps | Processing Time ms | Bit Rate Mbps | Processing Time ms | Bit Rate Mbps | Processing Time ms | Bit Rate Mbps |
| Mixed1 | 13.6 | 63 | 24.3 | 60 | 7.8 | 68 | 8.7 | 68 |
| Mixed2 | 40.7 | 21 | 47.5 | 31 | 35.8 | 15 | 36.7 | 16 |
| FileDownload1 | 7.3 | 118 | 21.9 | 67 | 2.5 | 216 | 3.4 | 177 |
| FileDownload2 | 8.3 | 104 | 19.5 | 75 | 2.9 | 187 | 3.4 | 176 |
| RemoteControl1 | 12.7 | 68 | 24.9 | 58 | 8.8 | 61 | 9.3 | 64 |
| RemoteControl2 | 8.1 | 107 | 20.8 | 70 | 2.7 | 197 | 3.0 | 201 |

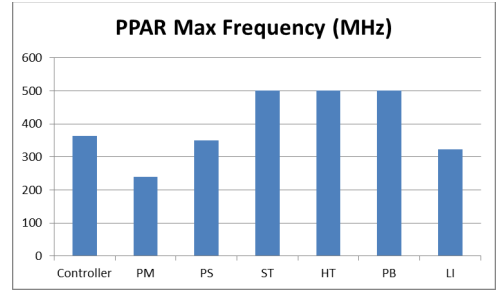

Fig. 4. Modules' speeds.

*C. Simulation Results*

The system was tested with 154kB patterns file containing a total of 2,500 patterns. The minimum length was 15 bytes and therefore, the maximum shift was 14 bytes. The simulation times taken by each trace file are shown in table 3. The bitrate decreased while processing the files doped with Set2, this is because Set2 has more patterns than other sets. In addition, the distribution of patterns in the same trace file affects the bitrate. This occurs because the implemented algorithm depends on skipping bytes according to shift table. If the distribution of patterns changed, the values of shifts retrieved from the shift table will differ according to the byte-pair that the PM stopped on. If the average of shifts decreased, the bit rate will decrease. This issue is clear in the bitrate of processing Mixed2 trace file as shown in Figure 5. Even though Set2 contains more patterns, but when it is mixed with Mixed2 trace file, it results in a better bitrate than mixing other sets with the same trace file. Nonetheless, the bitrate was calculated for the total size of the doped trace files.

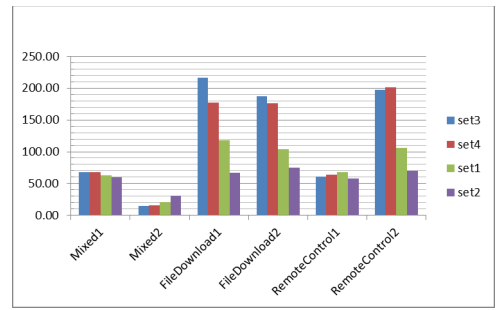

Fig. 5. Bitrate values.

When comparing the performance of our system to Snort, Spam filters' [34] or malicious websites [35] note that

ClamAV has over 1.5 million signatures as opposed to Snort's 3,000 signatures.

## VI. CONCLUSIONS

Using FPGAs to implement Wu-Manber matching algorithm has an advantage of increasing performance. But this is affected by several factors. Large patterns database decreases the speed of the system due to many factors. First the large number of patterns that the system should match with the traffic every time the system detects a pattern suffix. Second the average shift value from the shift table will decrease due to the increment of byte pairs that reside on patterns suffixes. Additionally, the bit rate of the system is variable due to the shift value retrieved from shift table which depends on the signatures as well as the traces.

## *Acknowledgment*

This work was supported by Zayed University Research Office, Research Incentive Grant # R17060.

## *References*

[1] E. M. El-Alfy, F. N. Al-Obeidat, A Multicriterion Fuzzy Classification Method with Greedy Attribute Selection for Anomaly-based Intrusion Detection, In Procedia Computer Science, Volume 34, 2014, pp. 55-62, ISSN 1877-0509, doi.org/10.1016/j.procs.2014.07.037.

[2] S. Aljawarneh, M. Aldwairi, M.B. Yassein, Anomaly-based intrusion detection system through feature selection analysis and building hybrid efficient model, (3)(2017).

[3] Y. Cho, S. Navab, W. Mangione-Smith, Specialized hardware for deep network packet filtering. 12th Int Conference on Field Programmable Logic and Application, (2002).

[4] M. Aldwairi, A. Abu-Dalo, M. Jarrah, "pattern matching for signature-based ids using Mapreduce framework and Myers algorithm", EURASIP J. on Info. Security, 1(9)(2017) 9. doi: 10.1186/s13635-017-0062-7.

[5] F. N. Al-Obeidat and E. S. M. El-Alfy, "Network Intrusion Detection Using Multi-Criteria PROAFTN Classification," 2014 International Conference on Information Science & Applications (ICISA), Seoul, 2014, pp. 1-5. doi: 10.1109/ICISA.2014.6847436.

[6] M. Aldwairi, K. Al-Khamaiseh, F. Alharbi, B. Shah. "Bloom filters optimized Wu-Manber for intrusion detection," Journal of Digital Forensics, Security and Law, 11(4)(2016) 5.

[7] M. Aldwairi, N. Ekailan, Hybrid pattern matching algorithm for intrusion detection systems. J. of Information Assurance and Security, 6(6)(2011) 512–521.

[8] M. Roesch, Snort-lightweight intrusion detection for networks. 13th Systems Administration Conference, Washington (1999).

[9] N. Horne, T. Kojm, T. Lamy, T. Madsen, D. D. Messemacker, T. Papszun, Trog, ClamAV User Manual. (2005).

[10] About ClamAV. (2012). URL:http://www.clamav.net/about. [Accessed 20/10/2015].

[11] ClamAV packages and ports. URL: http://www.clamav.net/download/packages/. [Accessed 25/10/ 2015].

[12] M. Aldwairi, R. Duwairi, W. Alqarqaz, A classification system for predicting RNA hairpin loops. Int Joint Conf on Bioinformatics, Systems Biology and Intelligent Computing, (2009) 109-115. doi: 10.1109/IJCBS.2009.123

[13] J. Korenek, P. Kopiersky, Traffic scanner-hardware accelerated intrusion detection system. TERENA Networking Conf. (2007).

[14] Z. K. Baker, V. K. Prasanna, A methodology for synthesis of efficient intrusion detection systems on FPGAs. 12th Annual IEEE Symposium on Field-Programmable Custom Computing Machines, (2004).

[15] D.E. Knuth, J.H. Morris, V.R. Pratt, Fast pattern matching in strings. SIAM J. Comput. 6(2)(1977) 323-350.

[16] R. S. Boyer, J. S. Moore, A fast string searching algorithm. ACM, (1977) 762-772.

[17] A. V. Aho, M. J. Corasick. Efficient string matching: an aid to bibliographic search. ACM, (1975) 333-340.

[18] S. Wu, U. Manber, A fast algorithm for multi-pattern searching. Tech. R. TR-94-17, Comp. Science, Univ of Arizona, (1994).

[19] M. Aldwairi, D. Alansari. Exscind: fast pattern matching for intrusion detection using exclusion and inclusion filters. Next Gen Web Services Practices (2011)(24-30). Salamanca, Spain. doi: 10.1109/NWeSP.2011.6088148

[20] S. Fide, A survey of string matching approaches in hardware. (2006)

[21] J. Tsung, L. Ho, G. G. F. Lemieux, PERG: A scalable FPGA-based pattern-matching engine with consolidated Bloomier filters. Int Conf. on Field-Programmable Technology, (2008) 73-80. Taipei.

[22] B. Zhang, X. Chen, X. Pan, Z. Wu, High concurrence Wu-Manber multiple patterns matching algorithm. (2009) Huangshan.

[23] Y. Mishina, K. Kojima, String matching on IDP: A string matching algorithm for vector processors and its implementation. ICCD (1993) Cambridge.

[24] Y. Fang, R. H. Katz, T. Lakshman, Gigabit rate packet pattern-matching using TCAM. 12th IEEE International Conference on Network Protocols, (2004) Washington.

[25] L. Tan, T. Sherwood, "A high throughput string matching architecture for intrusion detection and prevention," ACM Sigarch Computer, 33(2005) 112-122.

[26] M. Alicherry, M. Muthiprasanna, V. Kumar, High speed pattern matching for network IDS/IPS. IEEE International Conference on Network Protocols, (2006), Washington.

[27] M. Salour, X. Su, Dynamic two-layer signature-based IDS with unequal databases. Int. Conf. on Info. Tech, (2007) Washington.

[28] M. Kharbutli, M. Aldwairi, A. Mughrabi, Function and data parallelization of Wu-Manber pattern matching for intrusion detection systems. Net. Protocols and Alg., 4(3)(2012) 46-61.

[29] K. P. Themistoklis, S. K. Charalampos, G. M. Konstantinos, Parallel implementation of the Wu-Manber algorithm using the OpenCL framework. Artificial Intelligence Applications and Innovations, (2012) 576-583 Greece.

[30] G. M. Maier, Hardware pattern matching for network traffic analysis in gigabit environments. Diplomarbeit in Informatik, Technische Universität Munchen. (2007).

[31] N. I. Rafla, I. Gauba, A reconfigurable pattern matching hardware implementation using on-chip RAM-based FSM. 53rd IEEE Int. Midwest Symposium on Circuits and Systems, (2010) 49-52, WA.

[32] M. Aldwairi, M. Guled, M. Cassada, M. Pratt, D. Stevenson, P. Franzon. Switch architecture for optical burst switching networks. 1st workshop on Optical Burst Switching, (2003) Dallas.

[33] A. Orebaugh, G. Ramirez, J. Beale, J. Wright, Wireshark and Ethereal Network Protocol Analyzer Toolkit. (2007).

[34] M. Aldwairi and Y. Flaifel, "Baeza-Yates and Navarro approximate string matching for spam filtering", 2nd Int. Conf. on Innovative Computing Technology, (2012), Morocco, Sept 18-20.

[35] M. Aldwairi, R. Alsalman, MALURLs: Malicious URLs Classification System. Annual International Conference on Information Theory and Applications, (2011), Singapore, Feb 28. doi: 10.5176/978-981-08-8113-9_ITA2011-29